# Supervised Classification of Flow Cytometric Samples via the Joint Clustering and Matching (JCM) Procedure

Sharon X. Lee[1], Geoffrey J. McLachlan[1,*], Saumyadipta Pyne[2]

[1]Department of Mathematics, University of Queensland, St. Lucia, Queensland, 4072, Australia. [2]CR Rao Advanced Institute of Mathematics, Statistics and Computer Science, Hyderabad, India.
* E-mail: g.mclachlan@uq.edu.au

### Abstract

We consider the use of the Joint Clustering and Matching (JCM) procedure for the supervised classification of a flow cytometric sample with respect to a number of predefined classes of such samples. The JCM procedure has been proposed as a method for the unsupervised classification of cells within a sample into a number of clusters and in the case of multiple samples, the matching of these clusters across the samples. The two tasks of clustering and matching of the clusters are performed simultaneously within the JCM framework. In this paper, we consider the case where there is a number of distinct classes of samples whose class of origin is known, and the problem is to classify a new sample of unknown class of origin to one of these predefined classes. For example, the different classes might correspond to the types of a particular disease or to the various health outcomes of a patient subsequent to a course of treatment. We show and demonstrate on some real datasets how the JCM procedure can be used to carry out this supervised classification task. A mixture distribution is used to model the distribution of the expressions of a fixed set of markers for each cell in a sample with the components in the mixture model corresponding to the various populations of cells in the composition of the sample. For each class of samples, a class template is formed by the adoption of random-effects terms to model the inter-sample variation within a class. The classification of a new unclassified sample is undertaken by assigning the unclassified sample to the class that minimizes the Kullback-Leibler distance between its fitted mixture density and each class density provided by the class templates.

**Key terms** Flow cytometry, Sample of cells, Multiple samples, Clustering of cells, supervised classification of samples, Skew mixture models, EM algorithm

## 1 Introduction

Flow cytometry (FCM) is a powerful tool in clinical diagnosis of health disorders, in particular, immunological diseases. It offers rapid high-throughput measurements of multiple characteristics on every cell in a sample, capturing the fluorescence intensity of multiple fluorophore-conjugated reagents simultaneously. It is routinely used in clinical practices for the diagnosis of health disorders such as immunodeficiency diseases. FCM allows parallel measurements of multiple physical, chemical, and biological characteristics and properties of individual cells at a speed of several thousands of cells per second, thus generating rich and complex high-resolution data in a high-throughput manner. These parallel measurements of fluorescent intensities can be used to study the differential expression of different surface and intracellular proteins, enabling biologists to study a variety of biological processes at the cellular level.

A critical component in the pipeline of flow cytometry data analysis is the identification of cell populations from the multidimensional FCM dataset, currently performed manually by visually separating regions or gates of interest on a series of sequential bivariate projections of the data, a process known as *gating*. However, this approach has many problems and limitations, including variability between different analysts, non-standardization and non-reproducibility of results, an unrealistic assumption that biological relationships between the markers exist only in the projected low-dimensional space and, most importantly, non-scalability to high-dimensional analysis, especially involving a large number of samples. With the advancement of technology, modern day flow cytometers allow simultaneous measurements of many markers on millions of cells, with some latest revolutionary mass cytometers capable of extending this up to 100 simultaneous parameters [1, 2]. As the number of markers increases, the number of bivariate projections increase rapidly. This renders conventional manual analysis practically infeasible for unbiased FCM analysis. Due to this and the subjective and time-consuming nature of this approach, recent efforts have turned to the development of computational methods for the analysis of high-dimensional flow cytometry data to automate the gating process; see [3] for a recent account.

Among these methods, mixture models have been widely employed as the underlying mechanism for characterizing the heterogeneous cell populations [4, 5, 6, 7, 8, 9, 10, 11], taking advantage of the convenient and formal framework offered by a model-based approach to modelling these complex and multimodal datasets. Using this approach, the FCM data can be conceptualized as a mixture of populations each of which consists of cells with similar expressions, the distribution of which can be characterized by a parametric density. The task of cell-population identification then translates directly to the classical problem of multivariate model-based clustering. It is well known that data generated from flow cytometric studies are often asymmetrically distributed, multimodal, as well as having longer and/or heavier tails than normal. To accommodate this, several methods use a mixture of mixtures approach [7, 12] where a final cluster may consist of more than one mixture component, while some others adopt mixture models with skew distributions as components [4, 5, 7] to enable a single component distribution to correspond to a cluster.

In addition to cell-population identification, the classification of individual samples (for example, predicting the disease state of an unlabelled sample) presents another challenging task in the pipeline of flow cytometric data analysis. To this end, several techniques have been proposed recently in the literature, most of them based on extensions of algorithms for the cell-population identification mentioned above. In particular, many of these methods adopted a support vector machine (SVM) as their classifier, perhaps due to its simplicity, convenience, and computational efficiency. For example, the sample classifier of flowBin [3], SWIFT [12], ASPIRE [8], and flowPeaks [13] are all based on a SVM, trained on some selected features from their respective model fitted to the data. Typically, the cluster proportions (size) are used as the feature vector under these settings. An allocation of a sample to one of the predefined classes of samples is then done on the basis of this feature vector using a classifier formed from the available training data consisting of observations on the feature vector of known classification with respect to the classes. Sometimes different types of classifiers are used besides the SVM; for example, nearest neighbour in PBSC [3], and regularized regression in Citrus [14]. However, in all of these methods, only selected features are used in the post-hoc classification task. Hence their classifiers are based solely on the information in these features, ignoring a variety of other features of the underlying distributions such as their shapes or tails that may represent biologically interesting phenomena.

Here we present the JCM (Joint Clustering and Matching) procedure for the supervised classification of FCM samples with respect to a number of predefined classes. Unlike the above mentioned methods, JCM adopts a measure to quantify the similarity or dissimilarity between the fitted parametric models in terms of differences between the class templates representing the class densities of the markers being used. This approach provides a more complete assessment of the differences between the samples and the templates for different classes and so should lead to improved discrimination among the classes.

In constructing a template for the density of the markers for a given class of samples, JCM models the inter-sample variation in a class through the adoption of a random-effects model. With the exception of ASPIRE, the methods described above do not allow for possible inter-sample variations. They either explicitly or implicitly assume the cell populations to have the same underlying distribution across all samples. In particular, the location (or mode) and the shape of the distribution of these populations are taken to be the same. These assumptions are not realistic given that the cell populations typically vary significantly across different samples. These assumptions are relaxed within the JCM framework by modelling each sample as an instance of a class template, possibly transformed with a flexible amount of variation. The latter is governed by a random-effects model, thus allowing one to establish a direct parametric correspondence between the cell populations in a sample and their corresponding components in the mixture model for the class template. This formulation also means that the cell populations are automatically matched across different samples without any further post-processing.

Previously in [7], the effectiveness of JCM in clustering cells within a given a sample and aligning the cell populations across a batch of samples had been illustrated in three experiments, where JCM provided excellent results. In brief, JCM was applied to analyze two complex multi-class and multi-sample experiments, involving multiple time-points and multiple panels. In the second experiment, which is also examined here (but reformulated as a classification problem), JCM identified a spatio-temporal signature of BCR signaling that improved the distinction between the two classes of patients previously reported in [15]. In the third illustration, it was demonstrated how JCM was able to identify cell populations in a batch of Diffuse Large B-Cell Lymphoma (DLBCL) samples from the FlowCAP-I challenge [3] that closely matched the gated populations identified by expert analysts. In this paper, we focus on the application of the JCM procedure in the subsequent stage of the flow cytometry pipeline.

## 2 Materials and Methods

We shall demonstrate the effectiveness of JCM in classifying samples from different classes using a benchmark dataset from the FlowCAP-II challenge. In addition, we shall compare the classification performance of JCM with some other available algorithms on the follicular lymphoma dataset previously analyzed in [15].

### 2.1 Acute Myeloid Leukemia (AML) dataset

As part of the FlowCAP-II Sample Classification challenge, the Acute Myeloid Leukemia (AML) dataset was split equally into a training set and a test set. Peripheral blood or bone marrow samples from a total of 359 patients were collected; 316 were healthy patients and 43 were diagnosed with AML [3]. Due to the large number of markers used, the sample collected from each patient was assayed into eight tubes for staining by different markers combinations, with five markers from each tube. In addition, the forward scatter (FS) and side scatter (SS) were also available for each tube, yielding a seven-dimensional dataset. This totalled to 2872 files for analysis. The first and last tube (that is, tubes 1 and 8) were controls and hence not used in our analysis. The data have been transformed logarithmically for the SS measurements and all fluorescent markers, while the FS measurements remained linear. All channels have also been scaled to the unit interval during preprocessing. The dataset is available publicly from http://flowrepository.org/id/FR-FCM-ZZYA. For training, half of the dataset along with their known labels (156 healthy patients and 23 AML patients) was used. The challenge called for the construction of a classifier to label the samples in the test set (180 patients).

## 2.2 B-cell receptor (BCR) dataset

The B-cell receptor (BCR) dataset contains flow cytometric measurements from 28 patients diagnosed with follicular lymphoma (FL). In brief, each sample was split into eight multiplexed panels for staining by F(ab')2 against the BCR heavy chain. For each panel, a pair of lineage markers (common to all panels) was used to label the B-cells, while another pair (different in all panels) of phospho-markers was used to measure BCR signalling characteristics, totalling 18 different markers across all panels. Measurements were recorded for all panels at basal (unstimulated) and again at 4 minutes after stimulation. Further technical details on this experiment and the processing of samples can be found in the Supplementary Methods of [15]. It was observed in [16] that the FL patients can be stratified into two classes that have distinctly different survival outcomes, which is linked to the presence or absence of a subpopulation of B-cells known as the Lymphoma Negative Prognostic (LNP) cells. For our illustration, the dataset is randomly partitioned into a training and test set with equal number of samples in each set. The task for the automated algorithms is to determine the class of FL for each patient, based on the training samples provided.

## 2.3 The JCM algorithm

The JCM methodology is based on a multi-level model-based clustering approach, where each sample is represented by a flexible finite mixture model. The latter are intrinsically linked to a class template through a random-effects model (REM). In brief, the JCM approach can be conceptualized as a two-level hierarchical model consisting of a lower sample-specific level and a higher batch-specific level. At the sample-specific level, each cell population in a sample is characterized by a parametric multivariate distribution. A sample with multiple populations can thus be viewed as having a mixture distribution. Note that each sample has its own mixture model, implying that all the component parameters are specific to that sample. It should be pointed out that this is significantly different to some of the other available approaches, such as SWIFT and HDPGMM, that require that all samples to share the same component parameters (except the mixing proportions). Our approach gives JCM more flexibility in handling inter-sample variations. At the batch-specific level, an entire batch of samples can be modelled by a parametric multivariate template that describes the overall characteristics of the batch. To construct a representative template, JCM assumes that each sample-specific mixture model can be effectively modelled as an instance of a batch-specific (class) template with subtle inter-sample variations. This template is taken to be representative of the class from which the batch of samples is assumed to have been drawn. The class template is constructed by adopting a random-effects approach where individual mixture models are linked to the batch mixture models through a flexible affine transformation. A further advantage of this approach is that each cluster in a sample is automatically registered with respect to the corresponding cluster of the template. A third level can be envisaged when between-class comparisons in the case of multiple classes are made in a large cohort analysis. This additional higher level can be used to study intra-class relationships in situations where, for example, multiple disease types are present, individuals are measured at different time points, or multiple experiment conditions are analyzed. With the availability of an overall template for each class, classification of unlabelled samples can be easily performed by comparing its similarity with each class template. We describe here further technical details of each level.

### 2.3.1 Modelling of individual samples

As mentioned above, each sample is modelled by a finite mixture model. The JCM procedure provides two options for parametric densities, the default option using the multivariate skew $t$ (MST) distribution and an option using its symmetric version, the multivariate $t$-distribution. The MST distribution has additional parameters for handling heavy tails and skewness, rendering it well suited for modelling non-elliptical and asymmetrical clusters that are typical in flow cytometry data. Moreover, it formally encompasses the multivariate normal, $t$, and skew normal distributions as special or limiting cases.

To establish notation, let the $p$-dimensional vector $\boldsymbol{y}_{jk} \in \mathbb{R}^p$ contain the measurements (on $p$ markers) of cell $j$ in sample $k$, where $j = 1, \ldots, n_k$ and $k = 1, \ldots, K$. Here $n_k$ denotes the total number of cells in sample $k$, and $K$ denotes the total number of samples. We let $g$ denote the number of components in the mixture model. It is assumed that each cell in the $j$th sample belongs to one of the $g$ components. Then the density of $\boldsymbol{y}_{jk}$ can be expressed as

$$f(\boldsymbol{y}_{jk}; \boldsymbol{\Psi}_k) = \sum_{h=1}^{g} \pi_{hk} f(\boldsymbol{y}_{jk}; \boldsymbol{\theta}_{hk}), \tag{1}$$

where $\pi_{1k}, \ldots, \pi_{gk}$ denote the mixing proportions which are non-negative and sum to one. Here, $f(\boldsymbol{y}_{jk}; \boldsymbol{\theta}_{hk})$ denotes a component density with parameters specified by $\boldsymbol{\theta}_{hk}$ ($h = 1, \ldots, g$). The vector $\boldsymbol{\Psi}_k$ consists of all the unknown parameters of the mixture model, and is given by $\boldsymbol{\Psi}_k = (\pi_{1k}, \ldots, \pi_{g-1,k}, \boldsymbol{\theta}_{1k}^T, \ldots, \boldsymbol{\theta}_{gk}^T)^T$, where the superscript $T$ denotes vector or matrix transpose. In practice, the optimal value of $g$ can either be specified *a priori*, or inferred from the data using some information criterion such the Bayesian Information Criterion (BIC).

Regarding the choice of the component densities in (1), JCM has options: (i) the multivariate $t$-distribution and (ii) a multivariate skew $t$-distribution. In the former case, the $t$-distribution allows for heavier tails than the normal distribution, thus providing a more robust approach to traditional Gaussian mixture modelling (GMM). The $t$-component density is given by

$$t_p(\boldsymbol{y}_{jk}; \boldsymbol{\mu}_{jhk}, \boldsymbol{\Sigma}_{hk}, \nu_{hk}) = \frac{\Gamma(\frac{\nu_{hk}+p}{2})}{(\pi \nu_{hk})^{\frac{p}{2}} |\boldsymbol{\Sigma}_{hk}|^{\frac{1}{2}} \Gamma(\frac{\nu_{hk}}{2})} \left(1 + \frac{d_h(\boldsymbol{y}_{jk})}{\nu_{hk}}\right)^{-\frac{\nu_{hk}+p}{2}}, \tag{2}$$

where $\boldsymbol{\mu}_{jhk}$ is a (cell-specific) location parameter, $\boldsymbol{\Sigma}_{hk}$ is a positive-definite scale matrix, $\nu_{hk}$ is the degrees of freedom, $d(\boldsymbol{y}_{jk}) = (\boldsymbol{y}_{jk} - \boldsymbol{\mu}_{jhk})^T \boldsymbol{\Sigma}_{hk}^{-1} (\boldsymbol{y}_{jk} - \boldsymbol{\mu}_{jhk})$ denotes the squared Mahalanobis distance between $\boldsymbol{y}_{jk}$ and $\boldsymbol{\mu}_{jhk}$ (with $\boldsymbol{\Sigma}_{hk}$ as the scale matrix), and $\Gamma(\cdot)$ the Gamma function. The degrees of freedom $\nu_{hk}$ acts as a tuning parameter for regulating the thickness of the tails of the $t$-distribution, which can be inferred from the data. The second option for $f(\cdot)$ provided by JCM is a skew version of (2), with an additional vector $\boldsymbol{\delta}_{hk}$ consisting of $p$ skewness parameters for handling non-symmetric distributional shapes. In recent years, many different versions of the skew $t$-distribution have been proposed; see [17] for an overview on this topic. For our purposes, we shall adopt the popular version as proposed by [18]. Following the notation in [4], the skew $t$-density can be expressed as

$$\begin{aligned}
&ST_p(\boldsymbol{y}_{jk}; \boldsymbol{\mu}_{jhk}, \boldsymbol{\Sigma}_{hk}, \boldsymbol{\delta}_{hk}, \nu_{hk}) \\
&= 2t_p(\boldsymbol{y}_{jk}; \boldsymbol{\mu}_{jhk}, \boldsymbol{\Sigma}_{hk}, \nu_{hk}) \\
&\quad T_1\left(\boldsymbol{\delta}_{hk}^T \boldsymbol{\Omega}_{hk}^{-1}(\boldsymbol{y}_{jk} - \boldsymbol{\mu}_{hk})\sqrt{\frac{(\nu_{hk}+p)}{\nu_{hk} + d_h(\boldsymbol{y}_{jk})}}; 0, 1 - \boldsymbol{\delta}_{hk}^T \boldsymbol{\Omega}_{hk}^{-1} \boldsymbol{\delta}_{hk}, \nu_{hk} + p\right),
\end{aligned} \quad (3)$$

where $T_1$ denotes the (scalar) $t$-distribution function, and $\boldsymbol{\Omega}_{hk} = \boldsymbol{\Sigma}_{hk} + \boldsymbol{\delta}_{hk}\boldsymbol{\delta}_{hk}^T$. The estimation of the parameters of a mixture model can be carried out using the expectation-maximization (EM) algorithm. Specific details for finite mixtures of (2) can be found in [19], and in [4] for finite mixtures of (3).

### 2.3.2 Parametric batch-specific (class) template

To form a batch-specific (class) template, inter-sample variations are modelled through the introduction of random-effects (RE) terms that specify how sample-specific component distributions may vary from an overall representative mixture model (that is, the template model). More specifically, these RE terms govern how the component-location parameters $\boldsymbol{\mu}_{hk}$ relate to the batch location parameter $\boldsymbol{\mu}_h$. We proceed by assuming that each $\boldsymbol{\mu}_{hk}$ is an affine transformation of $\boldsymbol{\mu}_h$, that is,

$$\boldsymbol{\mu}_{hk} = \boldsymbol{a}_{hk} \circ \boldsymbol{\mu}_h + b_{hk}\mathbf{1}_p, \quad (4)$$

where $\circ$ denotes the Hadamard (or elementwise) product, $\mathbf{1}_p$ denotes a $p$-dimensional vector with all elements being one, and the scaling and translation RE terms are independent and given by

$$\begin{aligned}
\boldsymbol{a}_{hk} &\sim N_p(\mathbf{1}_p, \boldsymbol{\xi}_{1h}), \\
b_{hk} &\sim N_1(0, \xi_{2h}^2),
\end{aligned} \quad (5)$$

respectively. Estimation of the parameters of the mixture model with (4) and (2) or (3) can be implemented using the EM algorithm. The technical details of the fitting algorithm for our JCM model have been described in the Supplementary Information of [7].

### 2.3.3 Classification of new samples

In the case of multiple classes, a template can be formed for each class with JCM, using the available training data (that is, the samples of known origin with respect to the classes). As can be observed from (1) and (4), a JCM template is characterized parametrically as a mixture model similar to the sample-specific model fitted to each sample. This facilitates the quantitative comparison between different templates. Once the class templates are constructed, they can be quantitatively compared using a range of information-based measures, such as the Kullback-Leibler (KL) distance. In this approach, a new sample is fitted with a mixture model and its KL divergence from each of the templates is calculated. The new sample is then classified to the class associated with the smallest KL distance. We use a symmetric combination of the KL information.

Unlike ASPIRE and Citrus, which use only certain features calculated from the predicted clusters for each sample, the classification approach of JCM is based on the fitted densities for the class templates rather than their selected features. More specifically, these other methods calculate features such as the proportions and the median of some or all of the markers for the identified clusters as a simplified representation of each sample. A classifier is then built on these features using a support vector machine (SVM) (as in ASPIRE) or based on regularized regression (as in Citrus). As such, the accuracy of these classifiers can be quite sensitive to the choice of features, and is critically dependent on the distinctiveness of the selected features across the different classes. In contrast, JCM does not rely on any particular selected features from the fitted model, but rather uses the entire fit provided by the model by proceeding on the basis of the relative size of the class densities provided by the templates. When an unlabelled sample is presented to JCM, an independently fitted model of the sample is directly compared to the estimate of the parametric form of the template density for each class. The KL distance provides a measure of this difference between the estimated density of the unclassified sample and the fitted template density for each class.

# 3 Results

The performance of JCM is to be compared with some other methods in the classification of flow cytometry samples from two datasets. One of these two datasets is from the FlowCAP-II competition, while the second has been previously analyzed in [7, 15].

## 3.1 Performance evaluation

To evaluate the performance of JCM, we compute a number of measures as adopted in [3] namely sensitivity (or recall), specificity, accuracy, precision, and $F$-measure. Let $TP$ denote true positive, $TN$ denote true negative, $FP$ denote false positive, and $FN$ denote false negatives. Then the measures above are defined as:

$$sensitivity = \frac{TP}{TP+FN} \qquad (6)$$

$$specificity = \frac{TN}{TN+FP} \qquad (7)$$

$$accuracy = \frac{TP+TN}{TP+TN+FP+FN} \qquad (8)$$

$$precision = \frac{TP}{TP+FP} \qquad (9)$$

$$F\text{-measure} = 2 \times \frac{precision \times sensitivity}{precision + sensitivity} \qquad (10)$$

In addition, we report in the second last column of Tables 1 and 2 the adjusted Rand index (ARI) [20], a popular measure of cluster agreement in the model-based clustering literature. The performance of the JCM model is also evaluated by the area under receiver operating characteristic (ROC) curve (AUC), reported in the last column of Table 1.

## 3.2 Benchmark techniques

The performance of the JCM procedure is compared with some other available methods for sample classification, including HDPGMM [6], flowMatch[21], Citrus[14], and ASPIRE[8]. Like FLAME, JCM is based on finite mixtures of skew distributions. However, FLAME performs cluster alignment across samples in a post-hoc fashion using graph-matching techniques. In contrast, the multi-level modelling strategy of JCM allows for simultaneous clustering and matching cell populations across samples in a much more natural manner. Moreover, the availability of a parametric characterization of the class templates allows for the classification of unlabelled samples to be undertaken on the basis of differences between their densities and the class densities provided by the templates. Following FLAME, Azad et al. [21] proposed flowMatch for classifying samples based on templates. The flowMatch procedure treats each sample as a leaf node of a template tree, then performs agglomerative hierarchical clustering of the nodes to obtain a template tree, using a similarity measure based on the KL distance. However, with flowMatch, each node in the template tree is characterized by a single normal mixture model. Also, cluster matching is performed between two nodes before calculating their similarity measure. In contrast, JCM automatically matches the clusters across the samples and/or templates in the fitting process. The KL distance can be calculated directly on the fitted models under the JCM framework.

Recently, Cron et al. [6] proposed a hierarchical version (HDPGMM) of the Dirichlet process Gaussian mixture model (DPGMM) for the automatic alignment of cell populations across a batch of samples. Their model has an additional layer above the DPGMM models fitted to each sample, placing a hierarchical prior over the base distribution to link these individual DPGMMs. The HDPGMM model assumes that all samples share the same component parameters with the exception of the mixing proportions. In a more recent contribution, Dundar et al. [8] extended the HDPGMM model to allow for inter-sample random-effects using a much similar conceptual strategy to JCM. This model, known as ASPIRE, relaxes the requirement for component means to be the same across samples by assuming that they deviate from the corresponding template means probabilistically under a random-effects model. Using this approach, a global template is readily constructed without the need for post-hoc mode clustering as in HDPGMM.

Citrus [14] is another computational tool for sample classification and cell-population identification. Unlike the above mentioned methods, Citrus proceeds by a hierarchical clustering of cells selected from each sample, and the classification of samples is based on regression methods on selected features. For scalability, Citrus applies hierarchical clustering to aggregated data consisting of randomly sampled cells from each sample rather than the data pooled from all samples. This pooling approach removes the need for clustering individual samples, and automatically identifies local and global clusters, as well as matching them across different samples. From the hierarchical clustering of the pooled aggregated data, the proportion of each sample in each cluster is calculated. For each sample, a feature vector is formed consisting of the cluster proportions along with other features of the sample such as the median value of each marker.

## 3.3 Diagnosis of AML patients

We report in Table 1 the performance measures of JCM and other algorithms on the AML dataset. Since HDPGMM, flowPeaks, and SWIFT do not provide a strategy for sample classification, we follow a similar approach to [8]. For HDPGMM and flowPeaks, we extract the proportions of global clusters from each sample to form a feature vector to be used for training a SVM based on the labels from the training set. For SWIFT, a template or consensus model was computed based on the pooled samples. Subsequently, the cluster size for each sample were utilized as features for training a SVM. With ASPIRE, we adopted the traditional fully supervised classification setting as above, training a SVM on both the AML and normal cases in the training set.

Among these algorithms, JCM achieved the highest AUC value. Also, JCM along with Citrus achieved the highest $F$-measure value and ARI for this dataset. It can be observed from Table 1 that JCM and Citrus produced very similar results for the remaining performance measures. Indeed, both Citrus and JCM correctly predicted 179 of the 180 test-labels with Citrus incorrectly predicting an AML patient to be normal, whereas JCM mislabelled a normal patient to be AML. In a clinical situation, the latter case (higher sensitivity) may be more important (than higher specificity) when the test is used as a preliminary screening or early diagnosis, where the patients with a positive results can undergo further subsequent medical tests. As ASPIRE did not produce any results (terminated itself with unknown reasons before completing the fitting procedure), we do not report the results for ASPIRE in Table 1. However, it is noted in Table 3 in [8] that ASPIRE achieved an overall AUC of 98.9, although using a different partition into training and test sets.

Table 1: **Classification results by each algorithm on the AML dataset**

|          | Sensitivity | Specificity | Accuracy | Precision | $F$-measure | ARI  | AUC   |
|----------|-------------|-------------|----------|-----------|-------------|------|-------|
| JCM      | 1.00        | 0.99        | 0.99     | 0.95      | 0.97        | 0.97 | 0.997 |
| Citrus   | 0.95        | 1.00        | 0.99     | 1.00      | 0.97        | 0.97 | 0.975 |
| flowMatch| 0.85        | 1.00        | 0.98     | 1.00      | 0.92        | 0.89 | 0.925 |
| SWIFT    | 0.65        | 1.00        | 0.96     | 1.00      | 0.79        | 0.73 | 0.825 |
| flowPeaks| 0.45        | 1.00        | 0.94     | 0.62      | 0.62        | 0.55 | 0.725 |
| HDPGMM   | 0.45        | 1.00        | 0.94     | 0.62      | 0.62        | 0.55 | 0.725 |

The remaining two methods, HDPGMM and SWIFT, achieved perfect specificity, but performed poorly in terms of sensitivity. In other words, they have correctly identified all normal patients, but mislabelled a number of AML patients as being normal. This is also reflected in their poor $F$-measure, ARI, and AUC values.

## 3.4 Classification of BCR signalling cells

In [15], the FL patients were manually analyzed and classified into LNP$^-$ and LNP$^+$ classes. Here, we applied seven algorithms (JCM, ASPIRE, Citrus, flowMatch, flowPeaks, HDPGMM, and SWIFT) to this dataset. For training, half of the data was provided to each algorithm, together with their class labels. The BCR dataset presents a more challenging case for the algorithms, as the distinction between the classes is noticeably less pronounced than the AML dataset. In particular, it can be observed from Figure 1 that the distribution of the markers on the cells appears to be similar between an LNP$^-$ (Figure 1A) and LNP$^+$ (Figure 1B) sample. Perhaps the greatest difference occurs in the plot of the cells for the markers p-STAT5 and p.PLCg2 in Figures 1A3 and 1B3, where the shape of the distribution of the LNP$^-$ samples is asymmetrical compared to the LNP$^+$ sample.

The procedures JCM, ASPIRE, flowMatch, flowPeaks, HDPGMM, and SWIFT identified 5, 3, 4, 22, 16, and 1 clusters, respectively, with their templates. Their performance measures are reported in Table 2. It can be observed that JCM achieved the highest $F$-measure value of 0.71. Citrus and HDPGMM produced results that are quite similar to each other, yielding the same specificity and AUC. However, HDPGMM had a lower sensitivity, accuracy, and precision. With a zero sensitivity and precision, the performance of flowMatch, ASPIRE, flowPeaks, and SWIFT are relatively poor for this dataset. A closer look at the predicted classification labels given by ASPIRE, flowPeaks, and SWIFT reveals that they failed to discriminate between the two classes of patients, classifying all samples into one class.

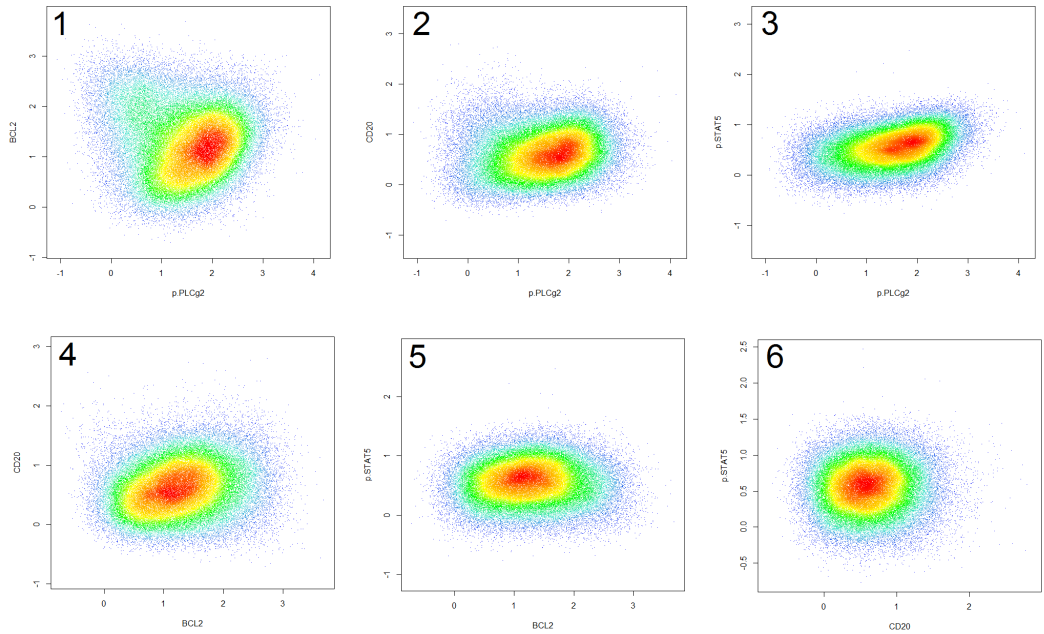

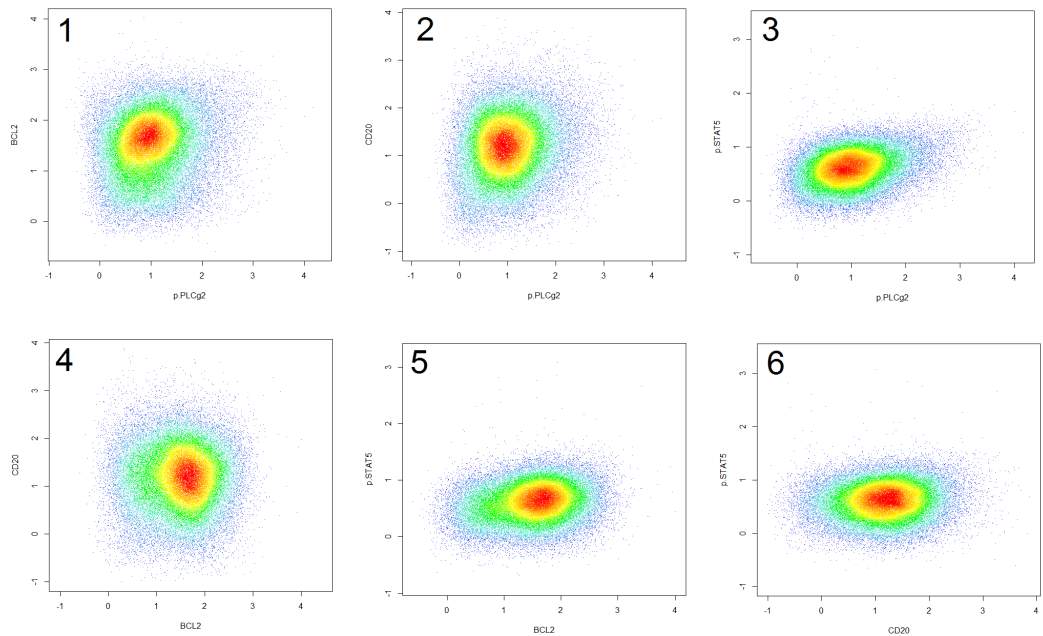

Figure 1: 2D scatter plots of markers from panel 4 for a sample from each of the LNP$^-$ (A) and a LNP$^+$ (B) in the BCR dataset.

Table 2: **Classification results by each algorithm on the BCR dataset**

|          | Sensitivity | Specificity | Accuracy | Precision | $F$-measure | AUC  |
|----------|-------------|-------------|----------|-----------|-------------|------|
| JCM      | 0.86        | 0.43        | 0.64     | 0.60      | 0.71        | 0.69 |
| Citrus   | 0.29        | 0.71        | 0.50     | 0.41      | 0.36        | 0.59 |
| HDPGMM   | 0.14        | 0.71        | 0.43     | 0.24      | 0.20        | 0.59 |
| flowMatch| 0.00        | 0.91        | 0.71     | 0.00      | 0.00        | 0.52 |
| ASPIRE   | 0.00        | 1.00        | 0.50     | 0.00      | 0.00        | 0.50 |
| flowPeaks| 0.00        | 1.00        | 0.50     | 0.00      | 0.00        | 0.50 |
| SWIFT    | 0.00        | 1.00        | 0.50     | 0.00      | 0.00        | 0.50 |

## 4 Modifications for a large number of markers

The largest number $p$ of markers considered simultaneously in the two examples above is seven. Our experience with the fitting of mixtures of restricted skew $t$ (rMST) distributions to multivariate data consisting of $p$ variables has shown that this mixture model is feasible in practice for $p$ as large as at least 20. It would suggest that the JCM procedure is therefore also feasible for 20 or so markers, although it does require more fitting time than with just the fitting of mixtures of rMST distributions since it also has additional random-effects terms in the model to construct the class template.

One way to reduce the dimensionality when there is a very large number $p$ of markers is to perform a principal component analysis (PCA) and work with the first $q$ principal components, where $q$ is chosen to be appropriately small.

Another way of proceeding in the case of very large $p$ is to use so-called matrix factorization. Let

$$\boldsymbol{X} = (\boldsymbol{y}_1, \ldots, \boldsymbol{y}_n)^T \tag{11}$$

be the $n \times p$ data matrix. Then $\boldsymbol{X}$ can be expressed via matrix factorization as

$$\boldsymbol{X} = \boldsymbol{X}_1 \boldsymbol{X}_2, \tag{12}$$

where $\boldsymbol{X}_1$ is a $n \times q$ matrix and $\boldsymbol{X}_2$ is a $q \times p$ matrix and where $q$ is chosen to be much smaller than $p$. For a specified value of $q$, the matrices $\boldsymbol{X}_1$ and $\boldsymbol{X}_2$ are chosen to minimize

$$\|\boldsymbol{X} - \boldsymbol{X}_1 \boldsymbol{X}_2\|^2, \tag{13}$$

where $\|\cdot\|$ is the Frobenius norm (the sum of squared elements of the matrix). With this factorization, dimension reduction is effected by replacing the data matrix $\boldsymbol{X}$ by the solution $\hat{\boldsymbol{X}}_1$ for the factor matrix $\boldsymbol{X}_1$; the $v$th column of $\hat{\boldsymbol{X}}_1$ gives the values of the $v$th metavariable for the $n$ observations in the sample. Thus the original $p$ variables are replaced by $q$ metavariables. When the elements of $\boldsymbol{X}$ are non-negative, we can restrict the elements of $\boldsymbol{X}_1$ and $\boldsymbol{X}_2$ to be non-negative. This approach is called non-negative matrix factorization (NMF) in the literature ([22]; [23]). We shall call the general approach where there are no constraints on $\boldsymbol{X}_1$ and $\boldsymbol{X}_2$, general matrix factorization (GMF). Nikulin and McLachlan [24] have developed a very fast approach to the general matrix factorization (13), using a gradient-based algorithm that is applicable to an arbitrary (differentiable) loss function.

The dimension-reduction procedures PCA, GMF, or NMF can be undertaken either before or after the deletion from each sample of cells with outlying expression values for some of the markers. These outliers can be deleted first by implementing JCM in the first instance for one or more subsets of the markers of manageable dimension. Also, one might wish to use just the front and side scatter measurements to identify outliers.

In applications of normal or $t$-mixture models to high-dimensional datasets outside of flow cytometry such as in the clustering of microarray gene-expression data, mixtures of factor analyzers (MFA), or mixtures of common factor analysers (MCFA) have been applied; see, for example, Baek et al. [25]. With this factor-analytic approach, the correlations between the variables in the component-covariance matrices are explained by the variables depending linearly on a small number $q$ of unobservable (latent) factors. The MCFA model differs from MFA in that it assumes a common matrix of factor loadings for the components in the mixture model before rotation of the factors to be white noise. The MCFA approach enables normal and $t$-mixture models to be fitted to dataset with dimension $p$ in the hundreds. It also allows the data to be displayed in low-dimensional spaces by plotting the data points in the space of the (estimated) factors. Unfortunately, the MCFA approach has not yet been extended to the case of component densities specified by the restricted skew normal or skew $t$-distribution so as to provide asymmetric clusters.

Concerning skew factor-analytic models for which there is methodology available for their implementation, there has been some work done on mixtures of generalized hyperbolic skew $t$ (GHST) factor analyzers. But it should be noted that the GHST distribution is different from the restricted skew $t$-distribution. It does not have heavy tails in all directions and it tends to the normal distribution and not the skew normal as the degrees of freedom become large. Lin et al. [26] have shown how mixtures of skew normal factor analyzers (MSNFA) can be implemented. This model can be used to cluster cells in high-dimensional samples in flow cytometry provided the outliers are first removed since the skew normal distribution like its symmetric counterpart is not robust to outliers.

## 5 Discussion

We have considered the JCM procedure for the clustering of cells within a sample and the unsupervised and supervised classification of multiple samples of cells for multidimensional flow cytometric datasets. JCM addresses these two problems with a single multi-level framework. Firstly, JCM can perform cell-population identification and alignment across multiple samples in a fully automated manner. Secondly, the template approach of JCM provides a convenient way to classify new samples. For the former task, the effectiveness of JCM has been illustrated in [7] on three experiments, giving promising results. The focus of this paper is on the latter task of sample classification.

For the problem of supervised classification of a new, unlabelled sample to one of a number of predefined classes, a parametric template for each class is constructed by JCM. Unlike FLAME, the class template built by JCM is characterized parametrically. This facilitates the use of divergence measures such as the KL distance for quantitative comparison of two parametric distributions. The use of the KL distance in classifying samples thus uses a different approach compared to methods such as HDPGMM, Citrus, and ASPIRE in which the classifier is built based on a limited number of features derived from the fitted models. In contrast, JCM is based on the fitted densities for the class templates and the sample to be classified and not just a few features of these estimated densities such as the estimated mixing proportions and component/cluster means or medians of some of the markers.

The JCM procedure was compared to existing benchmark methods on two real FCM datasets. In the AML classification challenge, JCM achieved an almost perfect AUC, the highest among the six methods (Citrus, flowMatch, SWIFT, flowPeaks and HDPGMM) con-sidered. The JCM procedure was also able to correctly identify all AML samples, achieving a sensitivity of one, while the sensitivity of none of the other five methods exceeded 0.95. On the BCR dataset, JCM performed best on the basis of both the $F$-measure and AUC.

# 6 Acknowledgement

The work of SXL and GJM is supported by a grant from the Australian Research Council. SP is supported by a Ramalingaswami Fellowship of DBT and grant from MoS&PI and DST CMS (Project No. SR/SA/MS:516/07 dated 21/04/2008).